\documentclass[12pt]{article}

\renewcommand{\d}{{\rm d}}
\newcommand{\bl}{\mbox{\boldmath$l$}}
\newcommand{\be}{\mbox{\boldmath$e$}}
\newcommand{\M}{{\cal M}}
\newcommand{\R}{{\cal R}}
\newcommand{\bT}{\overline{T}}
\newcommand{\invM}{\M^{-1}}

\begin{document}

\title{On the discrete Christoffel symbols
}

\author{V.M. Khatsymovsky \\
 {\em Budker Institute of Nuclear Physics} \\ {\em of Siberian Branch Russian Academy of Sciences} \\ {\em
 Novosibirsk,
 630090,
 Russia}
\\ {\em E-mail address: khatsym@gmail.com}}
\date{}
\maketitle
\begin{abstract}
The piecewise flat spacetime is equipped with a set of edge lengths and vertex coordinates. This defines a piecewise affine coordinate system and a piecewise affine metric in it, the discrete analogue of the unique torsion-free metric-compatible affine connection or of the Levi-Civita connection (or of the standard expression of the Christoffel symbols in terms of metric) mentioned in the literature, and, substituting this into the affine-connection form of the Regge action of our previous work, we get a second order form of the action. This can be expanded over metric variations from simplex to simplex. For a particular periodic simplicial structure and coordinates of the vertices, the leading order over metric variations is found to coincide with a certain finite difference form of the Hilbert-Einstein action.

\end{abstract}


MSC classes: 83C27; 53C05

keywords: Einstein theory of gravity; minisuperspace model; piecewise flat spacetime; Regge calculus; affine connection; Christoffel symbol; discrete connection

\section{Introduction}

The coordinateless description of general relativity (GR) proposed by Regge \cite{Regge} and later called Regge calculus was subsequently used in both the classical and quantum approaches to GR, see, eg, Ref \cite{RegWil}. This is a minisuperspace formulation of GR on a special (but everywhere dense in the configuration superspace of GR in an appropriate topology) class of the Riemannian spacetimes, namely, the piecewise flat ones. The latter means that the curvature lives on a set of zero 4-dimensional measure. Such a spacetime can be represented as a collection of the flat 4-dimensional tetrahedra or 4-simplices. The curvature is a $\delta$-function with support on the triangles or 2-simplices. The geometry is completely specified by the lengths of the edges or 1-simplices. Some functions on such a spacetime, including the GR action, can approximate their counterparts on the continuum spacetime with arbitrarily high accuracy \cite{Cheeger}. The discreteness of the set of variables, edge lengths, is important for its use in quantum theory, where the usual continuum formulation of GR is not formally renormalizable, see the works \cite{Ham,Ham1}. Also in these works, in particular, the subject of the connection in the discrete gravity was considered in detail.

The discrete analogs of the connection and curvature, orthogonal or arbitrary regular matrices, were introduced by Fr\"{o}hlich \cite{Fro}. In the so-called first order formalism, the connection is treated as an independent variable. In the framework of this formalism, the gravitational Wilson loop (a path-ordered exponential of the connection) was calculated in \cite{HamWil1,HamWil2}, where functional integration over the orthogonal connection as over an independent variable is performed for quantum averaging. There, the result of the paper \cite{CasDadMag} was used, in which the Regge calculus is formulated as a local theory of the Poincar\'{e} group with an action being a good approximation to the standard Regge action (of course, if the connection is replaced by its actual expression in terms of metric, as in the second order formalism considered now).

The problem of calculating the gravitational Wilson loop also arises in Causal Dynamical Triangulations theory of quantum gravity which uses a simplicial decomposition of spacetime and the Regge action as well, see \cite{cdt} and references therein. There, the connection orthogonal matrices on the three-dimensional faces (3-simplices) between the 4-simplices constituting the spacetime and the Wilson loop in terms of them are calculated depending on the configuration (geometry); quantum averaging is performed by summing over the configurations.

The GR action $\int R \sqrt{g} \d^4 x$ on the piecewise flat spacetime or the Regge action is a sum over the triangles $\sigma^2$,
\begin{equation}\label{SRegge}                                              
S = 2 \sum_{\sigma^2}{\alpha (\sigma^2) A (\sigma^2)},
\end{equation}

\noindent where $\alpha (\sigma^2)$ is the defect angle characterizing the curvature ($2 \pi$ minus the sum of the dihedral angles between the 3-simplices $\sigma^3 \supset \sigma^2$), $A ( \sigma^2 )$ is the area of $\sigma^2$. We consider the Euclidean metric signature $(+, +, +, +)$ for simplicity.

The defect angle refers to all the 3- or 4-simplices meeting at $\sigma^2$ as a whole, without division into certain contributions of individual simplices. At the same time, in the continuum GR action, $R$ is constructed of the Christoffel symbols $\Gamma^\lambda_{\mu \nu} ( \| g_{\lambda \mu} \| )$, which can be heuristically considered as corresponding to contributions into the curvature from certain directions or coordinate planes. To make the discrete expression more visual and similar to the continuum formula, we can construct the Regge action from matrices of a discrete connection. This requires a metric. To provide this, we should specify, in addition to the lengths for the edges $\sigma^1$, also the coordinates for the vertices $\sigma^0$. This allows us to define a piecewise affine metric that is constant in the 4-simplices. The final answer should not depend on the coordinates, but their knowledge allows us to distribute contributions to the curvature from the 3-simplices in a convenient way.

The problem of finding a specific discrete connection was addressed in the literature. It was noted in Ref \cite{HamWil} that the matrix which describes the parallel transport of any vector from one simplex to another neighboring one is related to the path-ordered exponential of the integral of the connection $\Gamma^\lambda_{\mu \nu}$. In Ref \cite{Drummond}, the connection field $\Gamma^\alpha_{\beta \gamma}$ on the simplicial manifold was obtained in some cases as having a certain (regularized) delta-function form with support on the 3-faces. There, a discrete torsion contribution from the connection was of interest. The Christoffel choice of connection was considered in Ref \cite{Barrett}. A parallel transport and a simplicial analogue of the covariant derivative of a vector was considered in Ref \cite{Kheyfets}. A discrete affine connection was considered in Ref \cite{Gronwald}. In our paper \cite{our}, using the discrete analogs of the connection and curvature introduced by Fr\"{o}hlich \cite{Fro}, we consider connection formulations of the Regge action (full orthogonal and self-dual ones) classically equivalent to the action (\ref{SRegge}) on the equations of motion for the connection. Upon a certain modification of such a formulation and the introduction of a piecewise affine coordinate frame on a simplicial manifold specified by attributing coordinates to the vertices, we get an exact representation of the action in terms of affine connection matrices (from GL(4,R)) \cite{our1}. There, we considered these matrices independent matrix variables; now we would like to fix them as a discrete analogue of the unique torsion-free metric-compatible (Levi-Civita) affine connection $\Gamma^\lambda_{\mu \nu}$.

Namely, in the present paper, we first consider general expressions for the connection and curvature matrices, mention some of their properties and write out the exact form of the action in terms of the metric. Then we specify this for a particular simplicial structure (the periodic one earlier used, eg, in Ref \cite{RocWil}) and coordinates of the vertices, consider the leading order over variations of the metric from simplex to simplex, give some examples of the connection and curvature matrices and show that the leading order of the action over metric variations (which just survives in the continuum limit) is a certain finite difference form of the Einstein action.

\section{The discrete connection and action in terms of the metric}

Consider a 3-face $\sigma^3$ shared by two 4-simplices whose vertices complementary to $\sigma^3$ are denoted $P$ and $Q$. These 4-simplices will be denoted as $\sigma^3P$ and $\sigma^3Q$. Some coordinates are assigned to the vertices, and we have a piecewise constant metric equal to $g_{\lambda \mu} (\sigma^3P )$ and $g_{\lambda \mu} (\sigma^3Q )$ in these 4-simplices, and a triplet of vectors $l^\lambda_i$, $i = 1, 2, 3$, defining the $\sigma^3$ plane. For $l^\lambda_i$, differences of the affine coordinates $x^\lambda_{\sigma^0}$ of the vertices $\sigma^0$ of $\sigma^3$ can be taken,
\begin{equation}\label{li=}                                                 
l^\lambda_i = x^\lambda_i - x^\lambda_O,
\end{equation}

\noindent if the vertices of $\sigma^3$ are $\sigma^0 = 1, 2, 3, O$.

We note some points regarding the introduction of the piecewise affine world coordinate system \cite{our1}. We freely set (in a non-degenerate way) the coordinates of all the vertices $\sigma^0$ at once. Inside each $\sigma^4$, the coordinates of its five vertices specify (the patch of) an affine coordinate system. The unambiguity, by construction, of the coordinates of any vertex, regardless of the simplex containing it, in the affine system of which these coordinates are set, guarantees the unambiguity of the coordinates when continuing from 4-simplices to 3-dimensional faces between them. In overall, these patches constitute a single piecewise affine coordinate system, an analogue of the continuum curvilinear world coordinate system. The edge lengths or, equivalently, a piecewise constant metric subject to the natural unambiguity conditions on the 3-faces (eq (\ref{lldg=0}) below) are given. Covering with a single coordinate system is a convenience as compared to the collection of the local frames of reference in the Euclidean frame formalism, but this is at the price of the less symmetric nature of the affine system, which is generally skew-angled and has different scales along the coordinate axes, in contrast to the Euclidean system. Arbitrary changes of the coordinates of the vertices (the edge lengths are fixed) are analogs of diffeomorphisms. Changes of the coordinates of the vertices generate GL(4,R) transformations in the 4-simplices, and the connection matrices are elements of GL(4,R) as well.

Let us define the result of the parallel transport of a vector from $\sigma^3P$ to $\sigma^3Q$ across the 3-face $\sigma^3$. For that, we transport as basis vectors $\bl_i$, $i = 1, 2, 3$ and the unit normal to $\sigma^3$,
\begin{equation}                                                            
n_\lambda = [\bl_1 \bl_2 \bl_3 ]_\lambda ( g^{\nu \rho} [\bl_1 \bl_2 \bl_3 ]_\nu [\bl_1 \bl_2 \bl_3 ]_\rho )^{- 1/ 2}, ~~ n^\lambda n_\lambda = 1, ~~ [\bl_1 \bl_2 \bl_3 ]_\rho \equiv \epsilon_{\lambda \mu \nu \rho} l^\lambda_1 l^\mu_2 l^\nu_3 ,
\end{equation}

\noindent $\epsilon_{\lambda \mu \nu \rho}$ being the naive ($\pm 1$) antisymmetric tensor. Then we can determine the matrix $\M^\lambda_{\sigma^3 \mu} \in \mbox{GL(4,R)}$ performing the parallel transport of a (contravariant) vector from $\sigma^3P$ to $\sigma^3Q$ by the condition that it acts on $l^\mu_i$, $i = 1, 2, 3$, identically and transforms $n^\mu ( \sigma^3P ) \equiv n^\mu ( \| g_{\nu \rho} ( \sigma^3P ) \| )$ to $n^\mu ( \sigma^3Q ) \equiv n^\mu ( \| g_{\nu \rho} ( \sigma^3Q ) \| )$,
\begin{eqnarray}\label{M=}                                                  
& & \M_{\sigma^3} : \bl_i \to \bl_i, ~~~ n^\mu ( \sigma^3P ) \to n^\mu ( \sigma^3Q ); \nonumber \\ & &
\M^\lambda_{\sigma^3 \mu} = \delta^\lambda_\mu - [ n^\lambda ( \sigma^3P ) - n^\lambda ( \sigma^3Q ) ] n_\mu ( \sigma^3P ) \nonumber \\ & & = \delta^\lambda_\mu - \frac{\sqrt{g( \sigma^3P )} \delta ( \sqrt{g} g^{\lambda \sigma} ) [\bl_1 \bl_2 \bl_3 ]_\sigma }{(g g^{\nu \rho}) (\sigma^3 \# ) [\bl_1 \bl_2 \bl_3 ]_\nu [\bl_1 \bl_2 \bl_3 ]_\rho} [\bl_1 \bl_2 \bl_3 ]_\mu , \\ & & \sigma^3 \# = \sigma^3P \mbox{ or } \sigma^3Q , ~~~ \delta f ( \sigma^4 ) \equiv f ( \sigma^3P ) - f ( \sigma^3Q ). \nonumber
\end{eqnarray}

\noindent This takes into account the unambiguity of the metric induced on the 3-face $\sigma^3$, in particular,
\begin{equation}\label{ggP=ggQ}                                             
(g g^{\nu \rho}) (\sigma^3P ) [\bl_1 \bl_2 \bl_3 ]_\nu [\bl_1 \bl_2 \bl_3 ]_\rho = (g g^{\nu \rho}) (\sigma^3Q ) [\bl_1 \bl_2 \bl_3 ]_\nu [\bl_1 \bl_2 \bl_3 ]_\rho
\end{equation}

\noindent in the denominator in (\ref{M=}) ( $\propto$ the volume of $\sigma^3$ squared, $(6 V_{\sigma^3})^2$).

Interchanging $P$ and $Q$ in $\M_{\sigma^3}$ we get such a matrix describing the transport from $\sigma^3Q$ to $\sigma^3P$, which we denote as $\M_{\overline{\sigma^3}}$, where $\overline{\sigma^3}$ stands for the same $\sigma^3$, but in this sense with a reversed orientation. Of course,
\begin{equation}                                                            
\M_{\overline{\sigma^3}} \M_{\sigma^3} = 1,
\end{equation}

\noindent as it should be. It's easy to see using
\begin{equation}                                                            
[\bl_1 \bl_2 \bl_3 ]_\lambda \M^\lambda_{\sigma^3 \mu} = \sqrt{ \frac{g( \sigma^3P )}{ g( \sigma^3Q )}} [\bl_1 \bl_2 \bl_3 ]_\mu ,
\end{equation}

\noindent which is obtained by taking into account the condition (\ref{ggP=ggQ}).

Also the construction provides that $\M$ is a metric-compatible connection,
\begin{equation}\label{MMg=g}                                               
\M^\lambda_{\sigma^3 \nu} \M^\mu_{\sigma^3 \rho} g^{\nu \rho } ( \sigma^3P ) = g^{\lambda \mu } ( \sigma^3Q ) .
\end{equation}

\noindent By explicitly checking this with taking into account (\ref{ggP=ggQ}), we get the formula
\begin{eqnarray}\label{MMg}                                                 
\M^\lambda_{\sigma^3 \nu} \M^\mu_{\sigma^3 \rho} g^{\nu \rho } ( \sigma^3P ) = \left ( g^{\lambda \mu} - \frac{g^{\lambda \sigma } [\bl_1 \bl_2 \bl_3 ]_\sigma [\bl_1 \bl_2 \bl_3 ]_\tau g^{\tau \mu }}{g^{\nu \rho } [\bl_1 \bl_2 \bl_3 ]_\nu [\bl_1 \bl_2 \bl_3 ]_\rho } \right )_{g^{\lambda \mu } = g^{\lambda \mu } (\sigma^3 P) } \nonumber \\ + \left ( \frac{g^{\lambda \sigma } [\bl_1 \bl_2 \bl_3 ]_\sigma [\bl_1 \bl_2 \bl_3 ]_\tau g^{\tau \mu }}{g^{\nu \rho } [\bl_1 \bl_2 \bl_3 ]_\nu [\bl_1 \bl_2 \bl_3 ]_\rho } \right )_{g^{\lambda \mu } = g^{\lambda \mu } (\sigma^3 Q) } .
\end{eqnarray}

\noindent The matrix in the first parenthesis, in fact, depends only on the metric induced on $\sigma^3$. Indeed, in the coordinate axes $l^\lambda_I$, $I = 1, 2, 3, P$, three of which coincide with $l^\lambda_i$, $i = 1, 2, 3$, and the fourth is associated, say, with the vertex $P$,
\begin{equation}\label{lI=}                                                
l^\lambda_P = x^\lambda_P - x^\lambda_O , ~~ l^\lambda_I = l^\lambda_i \mbox{ at } I = i = 1, 2, 3; ~~ g^{\lambda \mu } = l^\lambda_I g^{I J } l^\mu_J , ~~ I, J = 1, 2, 3, P ,
\end{equation}

\noindent this matrix reads
\begin{equation}                                                           
l^\lambda_I [ g^{IJ} - g^{IP}( g^{P P} )^{-1} g^{PJ} ] l^\mu_J .
\end{equation}

\noindent The matrix in the square brackets here is a nondegenerate $ 3 \times 3 $ block $g^{ij} - g^{i 4} (g^{44} )^{-1} g^{4 j}$, $i, j = 1, 2, 3$, bordered with zeros at $I = P$ and/or $J = P$. This block is the inverse of $\| g_{k l} \|$, $k, l = 1, 2, 3$. This $g_{k l} = l^\nu_k g_{\nu \rho} l^\rho_l$ is just the metric induced on $\sigma^3$, which is unambiguous, regardless of whether from $\sigma^3P$ or $\sigma^3Q$ this face is reached,
\begin{equation}\label{lldg=0}                                             
l^\nu_k l^\rho_l \delta g_{\nu \rho} \equiv l^\nu_k l^\rho_l [ g_{\nu \rho} (\sigma^3 P) - g_{\nu \rho} (\sigma^3 Q) ] = 0 .
\end{equation}

\noindent (This has the natural geometrical sense that any edge length is the same being defined in all the 4-simplices containing this edge. Or else that the 4-simplices should coincide on their common faces.) Therefore,
\begin{equation}                                                           
g^{\lambda \mu} - \frac{g^{\lambda \sigma } [\bl_1 \bl_2 \bl_3 ]_\sigma [\bl_1 \bl_2 \bl_3 ]_\tau g^{\tau \mu }}{g^{\nu \rho } [\bl_1 \bl_2 \bl_3 ]_\nu [\bl_1 \bl_2 \bl_3 ]_\rho } = l^\lambda_i ( \| l^\nu_k g_{\nu \rho } l^{\rho}_l \|^{- 1} )^{i j} l^\mu_j
\end{equation}

\noindent in (\ref{MMg}) can equally be taken at $g^{\lambda \mu } = g^{\lambda \mu } (\sigma^3 Q)$, which gives $g^{\lambda \mu } (\sigma^3 Q)$ for (\ref{MMg}), the RHS of (\ref{MMg=g}). An important role of the metric unambiguity conditions (\ref{lldg=0}) is seen here.

Of course, the above also follows from the fact that, due to (\ref{M=}), $\M_{\sigma^3}$ transforms an orthonormal basis $e^\lambda_i, n^\lambda ( \sigma^3P )$ to an orthonormal one $e^\lambda_i, n^\lambda ( \sigma^3Q )$ ($e^\lambda_i, i = 1, 2, 3$, define the plane of $\sigma^3$). Thus, $\M_{\sigma^3}$ simply re-expresses the same orthonormal basis $e_a, a = 1, 2, 3, 4, e_4 = n$, between the affine frames of $\sigma^3P$ and $\sigma^3Q$ without an additional rotation, so \cite{Ham,Ham1} $\M^\lambda_{\sigma^3 \mu} = e^\lambda_a ( \sigma^3Q ) e^a_\mu ( \sigma^3P )$. This corresponds to the definition of the discrete Levi-Civita connection \cite{Fro}.

Besides that, it is natural that $\M_{\sigma^3}$ could be given by the path-ordered exponential \cite{Ham,Ham1}
\begin{equation}\label{Pexp}                                               
{\cal P} \exp \left ( \int_C \Gamma_\lambda \d x^\lambda \right ) , ~~~ ( \Gamma_\lambda )^\mu_{~ \nu} \equiv \Gamma^\mu_{\lambda \nu} ,
\end{equation}

\noindent of the integral $\int_C \Gamma_\lambda \d x^\lambda$ along a path $C$ passing through $\sigma^3$ from any internal point of $\sigma^3 P$ to any internal point of $\sigma^3 Q$. (Here, the stepwise metric field is implied to be a limit of a regularized smooth function.) This is easy to see for small variations of the metric when we take the linear over $\Gamma^\lambda_{\mu \nu}$ part of (\ref{Pexp}); see below (\ref{dA=GdxA}).

Thus, $\M^\lambda_{\sigma^3 \nu}$ is a true finite analogue of the unique torsion-free metric-compatible affine connection or of the Levi-Civita connection $\Gamma^\lambda_{\mu \nu}$.

Consider the holonomy matrix $\R_{\sigma^2}$ of $\M_{\sigma^3}$. For a triangle $\sigma^2$, we choose a starting 4-simplex $\sigma^4_1 \supset \sigma^2$ and go around $\sigma^2$ in a closed loop successively passing through the 4-simplices and 3-faces containing this triangle. Let the triangle $\sigma^2$ be formed by some edges $\sigma^1_{\rm a}$, $\sigma^1_{\rm b}$ (and sometimes will be denoted as $\sigma^1_{\rm a} \sigma^1_{\rm b}$), the 3-faces $\sigma^3_n \supset \sigma^2$ are formed by the edges $\sigma^1_{\rm a}$, $\sigma^1_{\rm b}$ and some $\sigma^1_n$, and the 4-simplices $\sigma^4_n$ between the pairs $\sigma^1_n$, $\sigma^1_{n- 1}$ are formed by $\sigma^1_{\rm a}$, $\sigma^1_{\rm b}$, $\sigma^1_{n- 1}$, $\sigma^1_n$ (and will be denoted as $\sigma^2 \sigma^1_{n - 1} \sigma^1_n$ or $\sigma^2 \sigma^1_n \sigma^1_{n - 1}$). The connection matrices are
\begin{eqnarray}\label{Mn=}                                                
\M^\lambda_{\sigma^3_n \mu} = \delta^\lambda_\mu - \frac{[(\sqrt{g} g^{\lambda \tau})( \sigma^2 \sigma^1_n \sigma^1_{n - 1} ) - (\sqrt{g} g^{\lambda \tau})( \sigma^2 \sigma^1_{n+1} \sigma^1_n )] [\bl_{\sigma^1_{\rm a}} \bl_{\sigma^1_{\rm b}} \bl_{\sigma^1_n}]_\tau}{(g g^{\nu \rho})( \sigma^2 \sigma^1_n \sigma^1_{n \pm 1}) [\bl_{\sigma^1_{\rm a}} \bl_{\sigma^1_{\rm b}} \bl_{\sigma^1_n}]_\nu [\bl_{\sigma^1_{\rm a}} \bl_{\sigma^1_{\rm b}} \bl_{\sigma^1_n}]_\rho} \cdot \nonumber \\ \cdot \sqrt{g ( \sigma^2 \sigma^1_n \sigma^1_{n - 1} ) } [\bl_{\sigma^1_{\rm a}} \bl_{\sigma^1_{\rm b}} \bl_{\sigma^1_n}]_\mu , ~~~ n = 1, \dots , N, ~~~ \sigma^1_{N+1} \equiv \sigma^1_1 .
\end{eqnarray}

\noindent The holonomy reads
\begin{equation}\label{R=M...M}                                            
\R_{\sigma^2} = \M_{\sigma^3_N} \dots \M_{\sigma^3_{n + 1}} \M_{\sigma^3_n} \dots \M_{\sigma^3_1} .
\end{equation}

\noindent In this formula, all $\M$ act in the same direction along the loop enclosing $\sigma^2$. Then these $\M$ can enter in the form $\M^{-1}$ in the formulas for other $\R_{\sigma^2}$. This $\R_{\sigma^2}$ acts in the starting 4-simplex $\sigma^4_1 = \sigma^2 \sigma^1_N \sigma^1_1$. Using (\ref{MMg=g}),
\begin{equation}                                                           
\M^\lambda_{\sigma^3_n \nu} \M^\mu_{\sigma^3_n \rho} g^{\nu \rho} (\sigma^4_n ) = g^{\lambda \mu} (\sigma^4_{n + 1} ),
\end{equation}

\noindent gives that this is a rotation,
\begin{equation}\label{RRg=g}                                              
\R^\lambda_{\sigma^2 \nu} \R^\mu_{\sigma^2 \rho} g^{\nu \rho} (\sigma^4_1 ) = g^{\lambda \mu} (\sigma^4_1 ).
\end{equation}

\noindent Since each $\M_{\sigma^3_n}$ acts identically on $\bl_{\sigma^1_{\rm a}}$, $\bl_{\sigma^1_{\rm b}}$, it is a rotation around $\sigma^2$, and it is the rotation by the defect angle $\alpha( \sigma^2 )$. Extracting this defect angle from the matrix $\R_{\sigma^2}$, we can write out $S$ (\ref{SRegge}) (as in our paper \cite{our1} with taking into account (\ref{RRg=g})),
\begin{equation}\label{S}                                                  
S = 2 \sum_{\sigma^2} A ( \sigma^2 ) \arcsin \left [ \frac{ \R^\lambda_{\sigma^2 \tau} g^{\tau \mu} ( \sigma^4_1 )}{ 4 A ( \sigma^2 )} l^\nu_{\sigma^1_{\rm a}} l^\rho_{\sigma^1_{\rm b}} \epsilon_{\lambda \mu \nu \rho} \sqrt{g ( \sigma^4_1 )} \right ] .
\end{equation}

\noindent Here the function arcsin defines the defect angle on a triangle $\sigma^2$.

The action is nonzero due to nonzero variations of the metric, $\delta g_{\lambda \mu}$, more exactly, variations of the variable $g^{\lambda \mu} \sqrt{g}$ from simplex to simplex, $\delta ( g^{\lambda \mu} \sqrt{g} )$, over which $\M$ are linear. If the skeleton structure and the vertex coordinates are symmetric enough, each $\sigma^3$ is accompanied by a $\tilde{ \sigma }^3$ symmetric to it with respect to their common face $\sigma^2$ with the same orientation relative to the $x^\lambda$ coordinate axes, but passable in the opposite direction along the loop enclosing $\sigma^2$. Then
\begin{equation}                                                           
\R_{\sigma^2} = \dots \M_{\tilde{\sigma}^3 } \dots \M_{\sigma^3} \dots , \mbox{ where } \M_{\tilde{\sigma}^3 } \approx \M^{-1}_{\sigma^3} ,
\end{equation}

\noindent if the metric varies slowly from simplex to simplex. Then the leading order of the action over metric variations (only this order survives in the continuum limit) is given by $\delta m$- plus $(m)^2$-terms ($\M = 1 + m$), or $\delta^2 g$- plus $(\delta g)^2$-terms.

\section{Specifying the description for a particular simplicial structure}

As a working example, we use the simplest periodic skeleton structure with the hypercubic cell divided by diagonals into 24 4-simplices (used, eg, in Ref. \cite{RocWil}). We introduce the notation for the simplices referred to a given 4-cube or to a given vertex $O$ as to the origin. Direct the $x^\lambda$ axes along the sides of the 4-cube. The vertices of each 4-simplex besides $O$ can be obtained by successive shifts along $x^\lambda$, $x^\mu$, $x^\nu$, $x^\rho$, and we denote this simplex $[\lambda \mu \nu \rho ]$ (an ordered sequence). The diagonal connecting $O$ with the end point of the shifts along $x^\lambda$, \dots , $x^\nu$ is denoted by $(\lambda \dots \nu)$ (an unordered sequence). In particular, $\lambda$ is the 4-cube edge along $x^\lambda$. We denote any simplex formed by shifts along the diagonals and edges, by the corresponding ordered sequence of these diagonals and edges. We will omit the brackets $()$ and $[]$ in the cases in which this causes no confusion. In overall, we have the following simplices at the vertex $O$.

1) 15 edges $\lambda$, $\lambda \mu$, $\lambda \mu \nu$, $1 2 3 4$

2) 50 2-simplices (triangles) $\lambda \mu$, $(\lambda \mu ) \nu$, $\lambda (\mu \nu )$, $\lambda (\mu \nu \rho )$, $(\lambda \mu \nu ) \rho$, $(\lambda \mu ) (\nu \rho )$

3) 60 3-simplices (tetrahedra) $\lambda \mu \nu$, $(\lambda \mu ) \nu \rho$, $\lambda ( \mu \nu ) \rho$, $\lambda \mu ( \nu \rho )$ (the last three types will be denoted $\d \nu \rho$, $\lambda \d \rho$, $\lambda \mu \d$, respectively, where "$\d$" means "diagonal")

4) 24 4-simplices $\lambda \mu \nu \rho$

The shift operator along the edge $(\lambda \dots \nu)$ is denoted by $T_{\lambda \dots \nu }$ ($= T_\lambda \dots T_\nu $), the Hermitian conjugated $\bT$ is for shifting in the opposite direction.

By choosing the initial tetrahedra and the directions of the action of the matrices $\M_{\sigma^3}$ in a certain way, the matrices $\R_{\sigma^2}$ can be written as follows.
\begin{eqnarray}\label{R(M)}                                               
\R_{41} & = & \invM_{413}(\bT_2\invM_{241})(\bT_{23}\invM_{\d41})
(\bT_3\M_{341})\M_{412}\M_{41\d},\nonumber\\ \R_{4(23)} & = &
\invM_{4\d1}\invM_{423}(\bT_1\M_{14\d}) \M_{432},\nonumber\\ \R_{23} & = &
\invM_{23\d}\invM_{231}(\bT_4\invM_{423})(\bT_{14}
\M_{\d23})(\bT_1\M_{123})\M_{234},\nonumber\\ \R_{2(43)} & = &
\invM_{2\d1}\invM_{234}(\bT_1\M_{12\d}) \M_{243},\nonumber\\ \R_{(24)3} & = &
\invM_{\d31}\invM_{243}(\bT_1\M_{1\d3}) \M_{423},\\ \R_{1(32)} & = &
\invM_{132}(\bT_4\invM_{41\d})\M_{123} \M_{1\d4},\nonumber\\ \R_{1(432)} & = &
\invM_{1\d4}\M_{12\d}\M_{1\d3} \M_{14\d}\invM_{1\d2}\invM_{13\d},\nonumber\\
\R_{(14)(32)} & = & \invM_{14\d}\M_{\d23}\M_{41\d} \invM_{\d32},\nonumber\\
\lefteqn{\parbox{10cm}{\ldots 2 cycle perm (1, 2, 3)\ldots,}}\nonumber\\
\R_{4(123)} & = & \M_{4\d3}\invM_{42\d}\M_{4\d1}
\invM_{43\d}\M_{4\d2}\invM_{41\d}.\nonumber
\end{eqnarray}

\noindent These equations define 25 matrices $\R$. The remaining 25 matrices are obtained by permuting groups of indices: if $\R_{(\lambda \dots \mu)(\nu \dots \rho)} = \prod \bT_{(\dots)} \M^{\pm 1}_{ \dots \lambda \dots \mu \nu \dots \rho \dots }$, then $\R^{-1}_{(\nu \dots \rho)(\lambda \dots \mu)} = \prod \bT_{(\dots)} \M^{\pm 1}_{ \dots \nu \dots \rho \lambda \dots \mu \dots }$. This defines the action (\ref{S}).

The construction is invariant with respect to an arbitrary change of the coordinates of the vertices, which is an analogue of a diffeomorphism in the continuum theory. To fix the (piecewise affine) frame, we choose the coordinates of the vertices to run over the fours of integers $(n_1, n_2, n_3, n_4)$. The components of the metric tensor which obey the 3-face metric unambiguity conditions (\ref{lldg=0}) are (here implicitly) certain combinations of invariants - the edge lengths squared.

\section{Small metric variations}

Now we expand over the metric variations and find the leading $\delta^2 g$- plus $(\delta g)^2$-terms in the action. It is convenient to write the matrices $\M$ in the form $\M^\lambda_{~ \mu} = \delta^\lambda_\mu + g^{\lambda \nu} m_{\nu \mu}$, that is, first we lower the contravariant index on $\M - 1$. It is convenient to temporarily pass locally to the coordinate axes $l^\lambda_I$ (\ref{li=}), (\ref{lI=}) so that
\begin{equation}                                                           
\M^I_{\sigma^3 J} ( = (\| l^\nu_k \|^{-1})^I_\lambda \M^\lambda_{\sigma^3 \mu} l^\mu_J ) = \delta^I_J - \frac{\delta (\sqrt{g } g^{I P})}{\sqrt{g(\sigma^3 P )} g^{P P}( \sigma^3 P ) } \delta^P_J.
\end{equation}

\noindent In the first order in $\delta g$, we can transfer the variation $\delta$ from one factor to another ($g_{I K}$) using the product differentiating rule,
\begin{eqnarray}                                                           
m_{\sigma^3 I J} \equiv g_{I K} ( \M^K_{\sigma^3 J} - \delta^K_J ) = - g_{I K} \frac{\delta (\sqrt{g } g^{K P})}{\sqrt{g} g^{P P}} \delta^P_J = \delta^P_J \cdot \left\{
\begin{array}{rl} \delta g_{i P}, & I = i , \\ { \displaystyle \frac{1}{2} } \delta g_{P P}, & I = P .
\end{array} \right.
\end{eqnarray}

\noindent Remind that $\delta g_{i j} = 0$ (\ref{lldg=0}). In the original coordinates, we find
\begin{equation}                                                           
m_{\sigma^3 \lambda \mu} \equiv \M_{\sigma^3 \lambda \mu} - g_{\lambda \mu} = \left ( l^\nu_P \delta g_{\nu \lambda} - \frac{1}{2} \frac{ [\bl_1 \bl_2 \bl_3 ]_\lambda }{ [\bl_1 \bl_2 \bl_3 \bl_P ] } l^\nu_P l^\rho_P \delta g_{\nu \rho} \right ) \frac{ [\bl_1 \bl_2 \bl_3 ]_\mu }{ [\bl_1 \bl_2 \bl_3 \bl_P ] } .
\end{equation}

Let us apply this to the 3-face denoted as 123 in the above example of a periodic structure and formed by the successive shifts of the origin $O$ along the axes $x^1$, $x^2$, $x^3$. Here we have the following vectors $\bl_I$,
\begin{equation}                                                           
l^\lambda_1 = \left ( \begin{array}{c} 1 \\ 0 \\ 0 \\ 0 \end{array} \right ) , l^\lambda_2 = \left ( \begin{array}{c} 1 \\ 1 \\ 0 \\ 0 \end{array} \right ) , l^\lambda_3 = \left ( \begin{array}{c} 1 \\ 1 \\ 1 \\ 0 \end{array} \right ) , l^\lambda_P = \left ( \begin{array}{c} 1 \\ 1 \\ 1 \\ 1 \end{array} \right ) \left ( \begin{array}{r} \mbox{ also } l^\lambda_Q = \\ ( \equiv x^\lambda_Q - x^\lambda_O ) \end{array} \left ( \begin{array}{r} 0 \\ 0 \\ 0 \\ - 1 \end{array} \right ) \right )
\end{equation}

\noindent and
\begin{equation}\label{m123}                                               
m_{123 \lambda \mu} = \left ( \begin{array}{rrrr} 0 & 0 & 0 & \delta_4 g_{1 4} \\ 0 & 0 & 0 & \delta_4 g_{2 4} \\ 0 & 0 & 0 & \delta_4 g_{3 4} \\ 0 & 0 & 0 & {\displaystyle \frac{1}{2}} \delta_4 g_{4 4} \end{array} \right ) .
\end{equation}

\noindent Here $\delta_4$ means the variation corresponding to passing across the hyperplane $x^4 = 0$ in the positive direction of $x^4$.

The diagonal 3-face $12\d$ is formed by the successive shifts along the axes $x^1$, $x^2$ and the diagonal axis $x^3 = x^4$. The induced metric on the 3-face is ($\d x^4 = \d x^3$)
\begin{equation}\label{ds^2_12d}                                           
\d s^2_{12 \d } = g_{a b} \d x^a \d x^b + 2 ( g_{a 3} + g_{a 4} ) \d x^a \d x^3 + ( g_{3 3} + 2 g_{3 4} + g_{4 4} ) ( \d x^3 )^2, ~~~ a, b = 1, 2 .
\end{equation}

\noindent It should be unambiguous when approaching this face from both sides (the relations (\ref{lldg=0}) for $\sigma^3 = 12 \d$). The vectors of the vertices are
\begin{equation}                                                           
l^\lambda_1 = \left ( \begin{array}{c} 1 \\ 0 \\ 0 \\ 0 \end{array} \right ) , l^\lambda_2 = \left ( \begin{array}{c} 1 \\ 1 \\ 0 \\ 0 \end{array} \right ) , l^\lambda_3 = \left ( \begin{array}{c} 1 \\ 1 \\ 1 \\ 1 \end{array} \right ) , l^\lambda_P = \left ( \begin{array}{c} 1 \\ 1 \\ 0 \\ 1 \end{array} \right ) \left ( \mbox{ and } l^\lambda_Q = \left ( \begin{array}{r} 1 \\ 1 \\ 1 \\ 0 \end{array} \right ) \right ) ,
\end{equation}

\noindent so that
\begin{equation}\label{m12d}                                               
m_{12 \d \lambda \mu} = \left ( \begin{array}{rrrr} 0 & 0 & - \delta_{4 - 3} g_{1 4} & \delta_{4 - 3} g_{1 4} \\ 0 & 0 & - \delta_{4 - 3} g_{2 4} & \delta_{4 - 3} g_{2 4} \\ 0 & 0 & {\displaystyle \frac{1}{2}} \delta_{4 - 3} g_{3 3} & - {\displaystyle \frac{1}{2}} \delta_{4 - 3} g_{3 3} \\ 0 & 0 & - {\displaystyle \frac{1}{2}} \delta_{4 - 3} g_{4 4} & {\displaystyle \frac{1}{2}} \delta_{4 - 3} g_{4 4} \end{array} \right ) .
\end{equation}

\noindent Here $\delta_{4 - 3} = - \delta_{3 - 4}$ means the variation corresponding to passing across $\sigma^3 = 12\d$ in the positive direction of $x^4$ (and in the negative direction of $x^3$). It is taken into account that the variation $\delta_{4 - 3}$ of the induced 3-face metric (\ref{ds^2_12d}) vanishes due to its unambiguity, for example, $\delta_{4 - 3} ( g_{3 3} + 2 g_{3 4} + g_{4 4} ) = 0$.

For small variations of the metric, one can obtain the same $\M$, by representing the stepwise metric field as a limit of a smooth function, from the continuum formula
\begin{equation}\label{dA=GdxA}                                            
\delta A^\lambda = - \int \Gamma^\lambda_{\mu \nu} \d x^\mu A^\nu = \left ( g^{\lambda i} \delta g_{i 4} + \frac{1}{2} g^{\lambda 4} \delta g_{4 4} \right ) A^4
\end{equation}

\noindent for a vector $A^\lambda$. (Here the factors $g^{\lambda \mu}$ which multiply $\delta g_{\nu \rho}$ can be considered independent on $x$ and equal to $g^{\lambda \mu} ( \sigma^3 P ) \approx g^{\lambda \mu} ( \sigma^3 Q )$ in this order, in contrast to the case of considerable variations of the metric.) Here $g_{\lambda \mu}$ (and only $g_{i 4}$, $g_{4 4}$) are supposed to depend only on $x^4$, in a stepwise manner, and if in the above formulas the initial 4-simplex $\sigma^3 P$ corresponds to a certain point $x^\lambda$, then the final one $\sigma^3 Q$ corresponds to the point $x^\lambda + \d x^\lambda$,
\begin{equation}                                                           
\delta f \to -\frac{\partial f}{\partial x^4} \d x^4
\end{equation}

\noindent for our $\delta$.

Let us illustrate forming the matrices $\R_{\sigma^2}$ by the example of one of the given above 50 matrices (\ref{R(M)}) on the linear over metric variations level ($\delta^2 g$-part),
\begin{equation}                                                           
\R_{12} = \invM_{12\d} \invM_{123} (\bT_4\invM_{412}) (\bT_{34}\M_{\d12}) (\bT_3\M_{312}) \M_{124}.
\end{equation}

\noindent In addition to the orientations of the planes $\sigma^3$ for which $\M_{\sigma^3}$ are given, (\ref{m123}) and (\ref{m12d}), there is also the plane 124 with
\begin{equation}\label{m124}                                               
m_{124 \lambda \mu} = \left ( \begin{array}{rrrr} 0 & 0 & \delta_3 g_{1 3} & 0 \\ 0 & 0 & \delta_3 g_{2 3} & 0 \\ 0 & 0 & {\displaystyle \frac{1}{2}} \delta_3 g_{3 3} & 0 \\ 0 & 0 & \delta_3 g_{4 3} & 0 \end{array} \right ) .
\end{equation}

\noindent The other three $\sigma^3$ lay in close planes, but are shifted along $x^3$, $x^4$ or $x^3 = x^4$ and passed in the opposite direction along the loop enclosing $\sigma^2 = 12$. This just leads to the $\delta m$ part of $\R$,
\begin{eqnarray}                                                           
r_{12 \lambda \mu} \equiv \R_{12 \lambda \mu} - g_{\lambda \mu} & = & ( m_{124} - \bT_4 m_{412} ) - ( m_{123} - \bT_3 m_{312} ) - ( m_{12 \d} - \bT_{3 4} m_{\d 12} ) \nonumber \\ & \equiv & \delta_4 m_{124} - \delta_3 m_{123} - \delta_{4 + 3} m_{12 \d} .
\end{eqnarray}

\noindent Here, $\bT_4 m_{412}$, $\bT_3 m_{312}$, $\bT_{3 4} m_{\d 12}$ have the same functional dependence on the metric on both sides of the 3-faces $\sigma^3$ as $m_{124}$, $m_{123}$, $m_{12 \d}$, respectively; a small difference is only in the values of the metric itself; $\delta_{4 + 3}$ means the difference operation when passing to that simplex that is at a larger $x^3 = x^4$. So we have
\begin{eqnarray}                                                           
r_{12 \lambda \mu} & = & \left ( \begin{array}{rrrr} 0 & 0 & \delta_{4 + 3} \delta_{4 - 3} g_{1 4} + \delta_4 \delta_3 g_{1 3} & - \delta_{4 + 3} \delta_{4 - 3} g_{1 4} - \delta_3 \delta_4 g_{1 4} \\ 0 & 0 & \delta_{4 + 3} \delta_{4 - 3} g_{2 4} + \delta_4 \delta_3 g_{2 3} & - \delta_{4 + 3} \delta_{4 - 3} g_{2 4} - \delta_3 \delta_4 g_{2 4} \\ 0 & 0 & - {\displaystyle \frac{1}{2}} \delta_{4 + 3} \delta_{4 - 3} g_{3 3} + {\displaystyle \frac{1}{2}} \delta_4 \delta_3 g_{3 3} & {\displaystyle \frac{1}{2}} \delta_{4 + 3} \delta_{4 - 3} g_{3 3} - \delta_3 \delta_4 g_{3 4} \\ 0 & 0 & {\displaystyle \frac{1}{2}} \delta_{4 + 3} \delta_{4 - 3} g_{4 4} + \delta_4 \delta_3 g_{4 3} & - {\displaystyle \frac{1}{2}} \delta_{4 + 3} \delta_{4 - 3} g_{4 4} - {\displaystyle \frac{1}{2}} \delta_3 \delta_4 g_{4 4} \end{array} \right ) \nonumber \\ & \propto & \left ( \begin{array}{rrrr} 0 & 0 & 0 & 0 \\ 0 & 0 & 0 & 0 \\ 0 & 0 & 0 & 1 \\ 0 & 0 & - 1 & 0 \end{array} \right ) .
\end{eqnarray}

\begin{figure}[h]
\unitlength 1pt
\begin{picture}(120,115)(-180,-55)
\put(-5,0){\line(-1,0){15}}
\put(-50,0){\line(-1,0){10}}
\put(0,-42){\line(0,-1){13}}
\put(0,-5){\line(0,-1){27}}
\put(50,0){\vector(1,0){10}}
\put(6,0){\line(1,0){26}}
\put(0,50){\vector(0,1){10}}
\put(0,4){\line(0,1){35}}
\put(5,55){$x^4$}
\put(58,-10){$x^3$}
\put(-5,-4){$12$}
\put(45,45){\line(1,1){10}}
\put(5,5){\line(1,1){32}}
\put(-5,-5){\line(-1,-1){28}}
\put(-43,-43){\line(-1,-1){12}}
\put(-8,40){$124$}
\put(33,38){12d}
\put(33,-4){123}
\put(-50,-3){$\overline{T}_3 312$}
\put(-62,-42){$\overline{T}_{3 4}{\rm d}12$}
\put(-15,-42){$\overline{T}_4 412$}
\put(9,24){1}
\put(-20,20){2}
\put(-32,-17){3}
\put(-12,-27){4}
\put(18,-22){5}
\put(32,12){6}

\end{picture}
\caption{The 4-simplices 1, \dots , 6 and 3-faces 123, \dots , $\bT_4 412$ around the triangle 12 in the 34-plane.}\label{R12}
\end{figure}
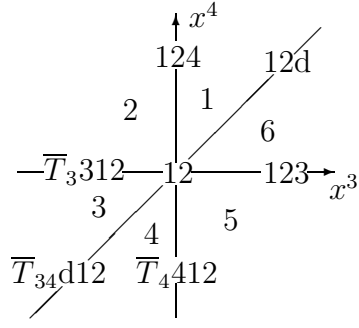

\noindent The last proportionality relation is due to strong cancellations in the expressions for the matrix elements, due to which $r_{12 \, 34} + r_{12 \, 43} = 0$ and $r_{12 \lambda \mu} = 0$ for any other pair $\lambda , \mu$ as a cyclic sum of variations on the 3-faces of certain components of the metric between the 4-simplices denoted as 1, 2, 3, 4, 5, 6 for brevity, see Fig.~\ref{R12}.
To see this, the unambiguity of the induced on the 3-faces metric should be taken into account. For example, consider $r_{12 \, 13}$. We use one of the unambiguity conditions $\delta_{4 - 3} ( g_{1 4} + g_{1 3} ) = 0$ of the induced metric on the diagonal 3-faces $1 2 \d$ and $\bT_{3 4} \d 1 2$ (\ref{lldg=0}). We use such a condition on the 3-faces 123 and $\bT_3 312$, $\delta_4 g_{1 3} = 0$. Thus, we form some identically zero expression $- \delta_{4 + 3} \delta_{4 - 3} ( g_{1 4} + g_{1 3} ) - \delta_3 \delta_4 g_{1 3}$, which we add to the element of interest and find that it is a cyclic sum of the variations $\delta g_{13}$,
\begin{eqnarray}                                                           
\delta_{4 + 3} \delta_{4 - 3} g_{1 4} + \delta_4 \delta_3 g_{1 3} = - \delta_{4 + 3} \delta_{4 - 3} g_{1 3} + \delta_4 \delta_3 g_{1 3} - \delta_3 \delta_4 g_{1 3} \nonumber \\ = [ - g_{1 3} ( 1 ) + g_{1 3} ( 6 ) + g_{1 3} ( 3 ) - g_{1 3} ( 4 ) ] + [ g_{1 3} ( 1 ) - g_{1 3} ( 2 ) \nonumber \\ - g_{1 3} ( 5 ) + g_{1 3} ( 4 ) ] + [ - g_{1 3} ( 6 ) + g_{1 3} ( 5 ) + g_{1 3} ( 2 ) - g_{1 3} ( 3 ) ] \equiv 0.
\end{eqnarray}

\section{The action in the leading order over metric variations}

Consider the sum of the contributions from the triangles $\sigma^2$ referred to a certain vertex/4-cube. The function $\arcsin{x}$ in (\ref{S}) can be taken as $x$ in this order. Since the matrices $\R_{\sigma^2}$ already have the required order ($\delta^2 g$- plus $(\delta g)^2$-terms), the metric functions $g^{\tau \mu} ( \sigma^4_1 ) \sqrt{ g ( \sigma^4_1 ) }$ multiplying them can be taken in the leading order the same for the considered group of $\R_{\sigma^2}$.

The matrix $\R_{\sigma^2}$ in the action $S$ is contracted with the edge vectors $\bl_{\sigma^1_{\rm a}}$, $\bl_{\sigma^1_{\rm b}}$. For the considered skeleton structure and coordinates of the vertices, any edge vector can be decomposed over the 4-cube edge vectors $\be_\lambda$ as
\begin{equation}                                                           
\bl_{\sigma^1} = \be_\lambda + \dots + \be_\mu \mbox{ for } \sigma^1 = ( \lambda \dots \mu ).
\end{equation}

\noindent (For the coordinates of the vertices used, $e^\lambda_\mu = \delta^\lambda_\mu$.) Then $l^{[ \nu}_{\sigma^1_{\rm a}} l^{\rho ]}_{\sigma^1_{\rm b}}$ can be decomposed into the "elementary" bivectors $e^{[ \nu}_\lambda e^{\rho ]}_\mu$. The latter can be considered as the bivector of the triangle $\lambda \mu$ or minus the bivector of the triangle $\mu \lambda$. In the action, the elementary bivector $e^{[ \nu}_\lambda e^{\rho ]}_\mu$ will enter contracted with a sum of matrices $\R_{\sigma^2}$. For example, the contribution of the bivector $e^{[ \nu}_2 e^{\rho ]}_3$ will be defined by the sum
\begin{equation}\label{R23+...}                                            
\R_{2 3} + \R_{2 (1 3)} + \R_{(2 1) 3} + \R_{(2 4) 3} + \R_{2 (4 3)} + \R_{(214) 3} + \R_{2 (143)} + \R_{(21) (43)} + \R_{(24) (13)} ,
\end{equation}

\noindent which can be viewed as that defining the triangle 23 contribution, minus the same one with 2 and 3 interchanged, which can be viewed as that defining the triangle 32 contribution. It turns out that the contribution of the triangle 32 in the considered order is the same as the contribution of the triangle 23.

Listed in the sum (\ref{R23+...}) are $\R$ on those $\sigma^2$ whose bivectors are obtained from the bivector of $\sigma^2 = 23$, $l^{[ \nu}_2 l^{\rho ]}_{(23)}$, by adding to $\bl_2$ and $\bl_{(23)}$ any possible edge vectors $\bl_{\sigma^1}$. Namely, there is a 3-prism in the direction $\bl_{\sigma^1}$ with the bases $\sigma^2 = 23$ and $T_{\sigma^1} 23$ and two internal triangles of interest $2 ( \sigma^1 3 )$ and $(2 \sigma^1 ) 3$. For example, in the direction 1, such internal triangles are $2 ( 1 3 )$ and $(2 1) 3$. In the direction 4, such triangles are $2 ( 4 3 )$ and $( 2 4 ) 3$. On the triangles $2 ( 1 3 )$ and $(2 1) 3$ as bases, there are 3-prisms in the direction 4 with the internal triangles $2(143)$, $(24)(13)$, $(21)(43)$, $(214)3$ of interest. This exhausts our list: the internal triangles of the 3-prisms in the direction 1 on $2(43)$ and $(24)3$ as bases and in the direction $(14)$ on $23$ as a base are already listed in the above way.

In the first order in $m$ ($\M = 1 + m$), the sum of interest (\ref{R23+...}) is a sum of certain variations $m_{\sigma^3} - m_{\tilde{\sigma}^3}$, where $\tilde{\sigma}^3$ and $\sigma^3$ are in close planes, but are not shifts of each other. Summing these contributions over triangles in the 3-prisms leads to the true finite difference expressions for the derivatives of the type $\Delta_\lambda m_{\sigma^3} \equiv (1 - \bT_\lambda ) m_{\sigma^3}$ with the function $m$ on the original $\sigma^3$ and shifted $\bT_\lambda \sigma^3$ 3-simplices. For example, the term $\bT_{14} m_{\d 23}$ from $\bT_{14} \M_{\d 23}$ in $\R_{23}$ turns out to appear in the sum (\ref{R23+...}) in the combination
\begin{eqnarray}                                                           
\R_{2 3} + \R_{2 (143)} + \R_{(214) 3} = \dots + (\bT_{14} m_{\d 23} - m_{23 \d }) + (m_{23 \d } - m_{2 \d 3}) \nonumber \\ + (m_{2 \d 3} - m_{\d 23}) + \dots = \dots + (\bT_{14} - 1)m_{\d 23} + \dots = \dots - \Delta_{14} m_{\d 23} + \dots
\end{eqnarray}

\noindent in the contribution from the 3-prism in the direction $(14)$ (made up of the tetrahedra $23 \d$, $2 \d 3$, $\d 23$) with the bases $\sigma^2 = 23$ and $T_{14} 23$ and the internal triangles $2(143)$ and $(214)3$. In overall, the linear in $m$ part of the sum (\ref{R23+...}) is
\begin{equation}                                                           
(1 - \bT_4 ) (m_{423} + m_{42 \d} - m_{4 \d 3}) - (1 - \bT_1 ) (m_{123} + m_{1 \d 3} + m_{12 \d}) - (1 - \bT_{14}) m_{\d 23} .
\end{equation}

\noindent Upon splitting $\Delta_{14} = \Delta_1 + \Delta_4 + O (\delta^2 )$, this reads
\begin{eqnarray}\label{D4M1-D1M4}                                          
\Delta_4 M_1 - \Delta_1 M_4 , \mbox{ where } M_1 = m_{423} + m_{42 \d } - m_{4 \d 3} - m_{\d 23} , \nonumber \\ M_4 = m_{123} + m_{12 \d } + m_{1 \d 3} + m_{\d 23} .
\end{eqnarray}

\noindent It is important that now in the form (\ref{D4M1-D1M4}) it is sufficient to know $m$ only in the leading order ($O ( \delta )$) to get the leading order in the final answer. Geometrically, say, $M_4$ means the parallel transport matrix along the 4-prism in the direction 4 with the base $\sigma^3 = 123$, but in the leading order, the matrix $M_4$ can be viewed as referring to six 4-prisms the bases of which are six permutations of $123$, that is, to the entire 4-cube.

Consider the $m^2$-terms in $\R$. We have expressions of the type
\begin{equation}                                                           
(1 + m_1)^{-1} (1 + \tilde{m}_2)^{-1} (1 + \tilde{m}_1 ) (1 + m_2 ) = 1 + \delta m_2 - \delta m_1 + [m_1 , m_2 ] + O( \delta^3 ) ,
\end{equation}

\noindent where $\delta m \equiv m - \tilde{m} = O (\delta^2 )$, and $m$ and their instances on close planes $\tilde{m}$ in the bilinear part of $\R$ are interchangeable in the leading order. For a larger number of $\M$ in $\R$ (six), all three pairwise combinations of $m$ are considered; for example,
\begin{equation}                                                           
\R_{23} = \dots + [m_{\d 23}, m_{123}] + [m_{\d 23}, m_{234}] + [m_{123}, m_{234}] + \dots .
\end{equation}

\noindent Such terms in the sum (\ref{R23+...}) are combined to give
\begin{equation}                                                           
\Delta_4 M_1 - \Delta_1 M_4 + [ M_4 , M_1 ] .
\end{equation}

\noindent In overall, such sums give the contribution
\begin{eqnarray}\label{sumR...}                                            
& & \frac{1}{2} \sum_{\sigma^2} \R^\lambda_{\sigma^2 \tau} g^{\tau \mu} \epsilon_{\lambda \mu \nu \rho} l^\nu_{\sigma^1_{\rm a}} l^\rho_{\sigma^1_{\rm b}} \sqrt{g} = \frac{1}{2} \sum_{\rm 4-cubes} \sum^4_{\sigma^1_1,\sigma^1_2 = 1} \left [ \R^\lambda_{\sigma^1_2 \sigma^1_1 \tau} + \dots + \R^\lambda_{(\sigma^1_2 \dots ) ( \sigma^1_1 \dots ) \tau} \right. \nonumber \\ & & \left. \phantom{\sum} + \dots \right ] g^{\tau \mu} \epsilon_{\lambda \mu \sigma^1_1 \sigma^1_2} \sqrt{g} = \sum_{\rm 4-cubes} ( \Delta_\lambda M_\mu - \Delta_\mu M_\lambda + [ M_\lambda , M_\mu ] )^{\lambda \mu} \sqrt{g}
\end{eqnarray}

\noindent to the action, where
\begin{eqnarray}                                                           
M_1 = m_{234} + m_{24 \d} - m_{34 \d} - m_{23 \d} , \nonumber \\
M_2 = m_{314} + m_{34 \d} - m_{14 \d} - m_{31 \d} , \nonumber \\
M_3 = m_{124} + m_{14 \d} - m_{24 \d} - m_{12 \d} , \nonumber \\
M_4 = m_{123} + m_{12 \d} + m_{31 \d} + m_{23 \d} .
\end{eqnarray}

\noindent (It turns out that to get the correct sign in $S = \int R \sqrt{g} \d^4 x$, one has to choose for the triangle $\sigma^2 = \rho \nu$ the following ordering of its edges: $\sigma^1_{\rm a} = (\rho \nu )$, $\sigma^1_{\rm b} = \rho$, which is just used in (\ref{sumR...}); $\epsilon_{1234} = + 1$.)

Let us examine $M_\lambda$ more closely. For $M_4$, the matrices $m_{123}$ (\ref{m123}) and $m_{12 \d}$  (\ref{m12d}) are already given, and we need $m_{31 \d}$ and $m_{23 \d}$,
\begin{equation}                                                           
m_{31 \d } \! = \! \left ( \begin{array}{rrrr} 0 & - \delta_{4 - 2} g_{1 4} & 0 & \delta_{4 - 2} g_{1 4} \\ 0 & {\displaystyle \frac{1}{2}} \delta_{4 - 2} g_{2 2} & 0 & - {\displaystyle \frac{1}{2}} \delta_{4 - 2} g_{2 2} \\ 0 & - \delta_{4 - 2} g_{3 4} & 0 & \delta_{4 - 2} g_{3 4} \\ 0 & - {\displaystyle \frac{1}{2}} \delta_{4 - 2} g_{4 4} & 0 & {\displaystyle \frac{1}{2}} \delta_{4 - 2} g_{4 4} \end{array} \right ) , m_{23 \d } \! = \! \left ( \begin{array}{rrrr} {\displaystyle \frac{1}{2}} \delta_{4 - 1} g_{1 1} & 0 & 0 & - {\displaystyle \frac{1}{2}} \delta_{4 - 1} g_{1 1} \\ - \delta_{4 - 1} g_{2 4} & 0 & 0 & \delta_{4 - 1} g_{2 4} \\ - \delta_{4 - 1} g_{3 4} & 0 & 0 & \delta_{4 - 1} g_{3 4} \\ - {\displaystyle \frac{1}{2}} \delta_{4 - 1} g_{4 4} & 0 & 0 & {\displaystyle \frac{1}{2}} \delta_{4 - 1} g_{4 4} \end{array} \right ) ,
\end{equation}

\noindent so that
\begin{equation}                                                           
M_4 = \left ( \begin{array}{rrrr} {\displaystyle \frac{1}{2}} \delta_{4 - 1} g_{1 1} & - \delta_{4 - 2} g_{1 4} & - \delta_{4 - 3} g_{1 4} & ( \delta_4 + \delta_{4 - 3} + \delta_{4 - 2} ) g_{1 4} - {\displaystyle \frac{1}{2}} \delta_{4 - 1} g_{1 1} \\ - \delta_{4 - 1} g_{2 4} & {\displaystyle \frac{1}{2}} \delta_{4 - 2} g_{2 2} & - \delta_{4 - 3} g_{2 4} & ( \delta_4 + \delta_{4 - 3} + \delta_{4 - 1} ) g_{2 4} - {\displaystyle \frac{1}{2}} \delta_{4 - 2} g_{2 2} \\ - \delta_{4 - 1} g_{3 4} & - \delta_{4 - 2} g_{3 4} & {\displaystyle \frac{1}{2}} \delta_{4 - 3} g_{3 3} & ( \delta_4 + \delta_{4 - 2} + \delta_{4 - 1} ) g_{3 4} - {\displaystyle \frac{1}{2}} \delta_{4 - 3} g_{3 3} \\ - {\displaystyle \frac{1}{2}} \delta_{4 - 1} g_{4 4} & - {\displaystyle \frac{1}{2}} \delta_{4 - 2} g_{4 4} & - {\displaystyle \frac{1}{2}} \delta_{4 - 3} g_{4 4} & {\displaystyle \frac{1}{2}} (\delta_4 + \delta_{4 - 3} + \delta_{4 - 2} + \delta_{4 - 1} ) g_{4 4} \end{array} \right ) .
\end{equation}

\noindent It can be shown that $( M_4 )_{\lambda \mu}$ is the finite difference form of $\Gamma_{\lambda , 4 \mu}$ obtained by substituting the derivatives $\partial_\lambda$ by the operators $\Delta_\lambda = 1 - \bT_\lambda $. Indeed, it is sufficient to consider the following five types of elements of the matrix $M$.

1) $(M_4)_{44} = \frac{1}{2} \Delta_4 g_{44}$, since $\Delta_4 \equiv \delta_{4 - 3} + \delta_{4 - 2} + \delta_{4 - 1} + \delta_4$

2) $( M_4 )_{14} = ( \delta_4 + \delta_{4 - 3} + \delta_{4 - 2} ) g_{1 4} - \frac{1}{2} \delta_{4 - 1} g_{1 1} = \Delta_4 g_{14} - \delta_{4 - 1} ( g_{14} + \frac{1}{2} g_{11} ) = \Delta_4 g_{14} - \frac{1}{2} \delta_{1 - 4} g_{44} = \Delta_4 g_{14} - \frac{1}{2} (\delta_{1 - 4} + \delta_{1 - 3} + \delta_{1 - 2} + \delta_1 ) g_{44} = \Delta_4 g_{14} - \frac{1}{2} \Delta_1 g_{44}$. It is taken into account that $\delta_{4 - 1} (g_{11} + 2 g_{14} + g_{44} )$, $\delta_{1 - 3} g_{44}$, $\delta_{1 - 2} g_{44}$, $\delta_{1 - 3} g_{44}$, $\delta_1 g_{44}$ are zero due to the 3-face metric unambiguity conditions (\ref{lldg=0}).

3) $(M_4)_{11} = \frac{1}{2} \delta_{4 - 1} g_{1 1} = \frac{1}{2} ( \delta_{4 - 3} + \delta_{4 - 2} + \delta_{4 - 1} + \delta_4 ) g_{1 1} = \frac{1}{2} \Delta_4 g_{11}$, since $\delta_{4 - 3} g_{1 1}$, $\delta_{4 - 2} g_{1 1}$, $\delta_4 g_{1 1}$ are zero due to the metric unambiguity.

4) $(M_4)_{41} = \frac{1}{2} \delta_{1 - 4} g_{4 4} = \frac{1}{2} ( \delta_{1 - 4} + \delta_{1 - 3} + \delta_{1 - 2} + \delta_1 ) g_{4 4} = \frac{1}{2} \Delta_1 g_{44}$, since $\delta_{1 - 3} g_{4 4}$, $\delta_{1 - 2} g_{4 4}$, $\delta_1 g_{4 4}$ are zero due to the metric unambiguity.

5) $(M_4)_{12} = \delta_{2 - 4} g_{1 4}$. Simply check the double target expression, $\Delta_2 g_{14} + \Delta_4 g_{12} - \Delta_1 g_{42} = ( \delta_{2 - 4} + \delta_{2 - 3} + \delta_{2 - 1} + \delta_2 ) g_{14} + ( \delta_{4 - 3} + \delta_{4 - 2} + \delta_{4 - 1} + \delta_4 ) g_{12} - ( \delta_{1 - 4} + \delta_{1 - 3} + \delta_{1 - 2} + \delta_1 ) g_{42} = ( \delta_{2 - 4} + \delta_{2 - 1} ) g_{14} + ( \delta_{4 - 2} + \delta_{4 - 1} ) g_{12} - ( \delta_{1 - 4} + \delta_{1 - 2} ) g_{42} = \delta_{2 - 4} ( g_{14} - g_{12} ) + \delta_{2 - 1} ( g_{14} + g_{24} ) + \delta_{4 - 1} ( g_{12} + g_{42} ) = 2 \delta_{2 - 4} g_{1 4}$, as required. Here $\delta_{2 - 3} g_{14}$, $\delta_2 g_{14}$, $\delta_{4 - 3} g_{12}$, $\delta_4 g_{12}$, $\delta_{1 - 3} g_{42}$, $\delta_1 g_{42}$, $\delta_{2 - 4} ( g_{14} + g_{12} )$, $\delta_{2 - 1} ( g_{14} + g_{24} )$, $\delta_{4 - 1} ( g_{12} + g_{42} )$ are zero due to the metric unambiguity.

Significant is that the number of the independent nonzero values $\delta_{\lambda - \mu} g_{\nu \rho}$ and $\delta_\lambda g_{\mu \nu}$ coincides with the number of the components $\Delta_\lambda g_{\mu \nu}$ per 4-cube/vertex. Indeed, the number of the operators $\delta_{\lambda - \mu} = - \delta_{\mu - \lambda}$ (or the number of the diagonal 3-planes) is $d \frac{d - 1}{2}$. The number of the operators $\delta_\lambda$ (or the number of the coordinate 3-planes) is $d$. For each 3-plane, there are $d$ independent nonzero metric variations $\delta g$. This gives $d^2 \frac{d + 1}{2}$ nonzero independent values $\delta g$, the same as the number of the components $\Delta_\lambda g_{\mu \nu}$ (40 for $d = 4$). The finite differences $\delta g$ and $\Delta g$ are expressible in terms of each other.

Thus, $(M_\mu )^\lambda_{ ~ \nu } \equiv M^\lambda_{\mu \nu}$ is the finite difference form of $\Gamma^\lambda_{\mu \nu}$, and the action (\ref{sumR...}) reads
\begin{eqnarray}\label{DM+MM}                                              
\hspace{-5mm}
\sum_{\rm 4-cubes} K^{\lambda \mu}_{~~~ \lambda \mu} \sqrt{g} , \mbox{ where } K^\lambda_{~ \, \mu \nu \rho} = \Delta_\nu M^\lambda_{\rho \mu} - \Delta_\rho M^\lambda_{\nu \mu} + M^\lambda_{\nu \sigma} M^\sigma_{\rho \mu} - M^\lambda_{\rho \sigma} M^\sigma_{\nu \mu} , \nonumber \\
M^\lambda_{\mu \nu} = \frac{1}{2} g^{\lambda \rho} (\Delta_\nu g_{\mu \rho} + \Delta_\mu g_{\rho \nu} - \Delta_\rho g_{\mu \nu}), ~~~ \Delta_\lambda = 1 - \bT_\lambda .
\end{eqnarray}

\noindent $K^\lambda_{~ \, \mu \nu \rho}$ is the finite difference form of the Riemannian tensor $R^\lambda_{~ \, \mu \nu \rho}$.

\section{Conclusion}

Our description of the system implies setting the edge lengths and coordinates of the vertices. This allows to define the piecewise constant metric, a finite analogue $\M$ (\ref{M=}), (\ref{Mn=}) of the unique torsion-free metric-compatible Levi-Civita affine connection (Christoffel symbols), the holonomy of this connection or curvature matrices $\R$ (\ref{R=M...M}) and the Regge action in terms of them (\ref{S}). The matrices $\M$ are linear over variations of the metric from simplex to simplex (or, precisely speaking, of the variable $g^{\lambda \mu} \sqrt{g}$).

Such a form of $\M$ means that there is no need to introduce any fixed background metric or lengths for the expansion over metric variations. (And the leading order of such an expansion just survives in the continuum limit.) The action resembles the continuum Einstein-Hilbert one constructed using the Christoffel symbols, linear in the derivatives of the metric.

For the simplest periodic skeleton structure with the cubic cell and the simplest appropriate choice of the piecewise affine frame (choosing the coordinates of the vertices to run over the fours of integers), we have obtained a certain finite difference form of the Einstein-Hilbert action when approaching the continuum limit, (\ref{DM+MM}).

We have obtained ten components of the metric tensor per 4-cube/vertex as dynamical degrees of freedom in this order. Only finite differences between 4-cubes (and not between neighboring 4-simplices) enter the action in this order. The action is weakly dependent on five of the fifteen edge lengths or their  combinations per 4-cube/vertex, provided that the metric variations remain actually small; we can take, eg, fixed values for them (as a kind of gauge conditions) or take into account subsequent orders, which govern their dynamics.

This agrees with Ref. \cite{RocWil}, where it was shown that the free theory of small fluctuations of the considered Regge skeleton about flat space gives the correct continuum limit. This paper gives ten dynamical degrees of freedom out of fifteen edge lengths per 4-cube/vertex, the other five decouple.

In subsequent orders over metric variations or in the exact approach, we could continue to use the piecewise constant metric as a variable. But generally the metric variations between 4-simplices enter the action in a non-degenerate way, and the metric unambiguity conditions on the 3-simplices (\ref{lldg=0}) should be taken into account. Therefore, it is more convenient to substitute the metric in terms of the edge lengths squared in this case (in each 4-simplex the metric is a linear combination of its edge lengths squared).

Then the role of choosing the coordinates of the vertices is in distributing the total curvature on a triangle among the (matrices $\M$ on the) tetrahedra containing this triangle. For example, in principle, we can choose the coordinates so that the matrices $\M$ be 1 for all but one of the tetrahedra containing the given triangle. An alternative is the above consideration when $\M_{\sigma^3}$ depends only on the parameters of the $\sigma^3$ environment (two $\sigma^4$ containing $\sigma^3$).

In addition to analyzing the effectiveness of the simplicial approximation to the continuum, an obvious case of using the considered formalism arises when we analyze the skeleton dynamics itself, and not just try to study the continuum dynamics approximately. The Regge skeleton equations seem to be much more difficult to solve analytically than their continuum counterpart. The expected finite difference form of the Einstein continuum equations, apparently, can be a reasonable intermediate version in complexity.

For example, we can consider discretizing a spherically symmetrical system. As a result, the symmetry will be broken, but the standard view of the situation is that the symmetry should be restored as a result of averaging over all possible kinds of Regge decompositions into the simplices. The case of the Schwarzschild and Reissner-Nordstrøm geometries was considered in Ref \cite{Wong}. It turned out to be expedient to use some (fixed) icosahedral decomposition (of three-dimensional space) into tetrahedra, since its symmetry is as close to spherical as possible. There, the elementary lengths obviously depend substantially on the radius.

But we would like to analyze the case when the background elementary lengths due to specific properties of the discrete path integral measure are loosely fixed dynamically at the level of a certain Planck scale value \cite{our2}. Here, the use of the above simplest periodic simplicial structure may be appropriate, since it respects some sufficient uniformity of the elementary length scale and at the same time allows simple averaging over the orientation (global) of the cubic axes. This allows us to use the above formalism and in general terms reduce the problem to the analysis of the finite difference form of the Einstein equations (in asymptotically Cartesian coordinates). The determining factor may be the requirement that at large distances (where the metric variations are small) the solution of interest should pass into a known continuum solution. It is expected that one of the properties of the solution (also analytically continued to non-integer coordinates) will be the absence of infinity in the central singularity. This can be illustrated by the Newtonian problem, in which the finite difference form of the Poisson equation for the gravitational potential with a $\delta$-function-like source gives a solution that is finite at the origin: it can be said that it is cut off at the elementary length scale. Of course, this requires a more detailed analysis.

\section*{Acknowledgments}

The present work was supported by the Ministry of Education and Science of the Russian Federation

\end{document}